\documentstyle[amsfonts,12pt,a4]{article}
\begin{document}
\begin{flushright}
quant-ph/9611012
\end{flushright}
\begin{flushleft}
{Modern Physics Letters A, Vol. 11, No. 19 (1996) 1563-1567}
\end{flushleft}

\begin{center}
\vspace{1 cm}

{\bf NEW FEATURES IN SUPERSYMMETRY BREAKDOWN IN QUANTUM MECHANICS}

\vspace{1 cm}Boris F. Samsonov \footnote{$^{}$ {email:
samsonov@phys.tsu.tomsk.su}}\vspace{1 cm}

Tomsk State University, 36 Lenin Ave. \\634050, Tomsk, Russia \\

\vspace{2 cm}
\end{center}

\begin{abstract}
The supersymmetric quantum mechanical model based on higher-derivative
supercharge operators possessing unbroken supersymmetry and discrete
energies below the vacuum state energy is described. As an example harmonic
oscillator potential is considered
\end{abstract}


{\bf 1}. Ideas of the supersymmetry have appeared in physics for the first
time in the quantum field theory for unifying the interactions of a
different nature \cite{r1}. In a supersymmetric theory the supersymmetry can
be either exact or spontaneously broken. The supersymmetric quantum
mechanics has been introduced \cite{wit} to illustrate the problems of the
supersymmetry breakdown in supersymmetric quantum field theories. For this
purpose the Witten criterion based on the Witten index \cite{wit} has been
elaborated. In the case of the broken supersymmetry the entire spectrum of
the super-Hamiltonian is twofold degenerate and in the case of the exact one
its vacuum state is nondegenerate. In the first case the supercharge
operators map the two states corresponding to the vacuum energy (zero
energy) one into another and in the second one the vacuum state is
annihilated by both supercharges. (See for example a recent survey \cite
{Coop}.)

Recently higher-order derivative extension of the supersymmetric quantum
mechanics has been elaborated \cite{andr}. In this approach supercharges are
constructed in terms of the higher-derivative differential operators and the
corresponding superalgebra is polynomial in the Hamiltonian. This model
exhibits a number of unusual properties \cite{anpr}. In particular, the
Witten criterion of spontaneous supersymmetry breaking is no longer
applicable \cite{andr}.

We now want to describe an unusual property of such models in terms of
supersymmetry breakdown which has not been described earlier. In our case
the state which is nondegenerate and annihilated by both mutually conjugated
supercharges is situated in the middle of the discrete spectrum of a
super-Hamiltonian. It follows that if one associates with this state the
zero energy value, the underlying energies should take negative values. This
situation will not occur in the conventional supersymmetric quantum
mechanics \cite{Coop} and we can claim that our higher-derivative model
exhibits at once the properties of the models with both exact and
spontaneously broken supersymmetry.

{\bf 2}.The higher-derivative supersymmetry in quantum mechanics \cite{andr}
is closely related to the higher-derivative Darboux transformation \cite
{anpr}, \cite{bstmf}, \cite{atmf}. This transformation, denoted here as $%
L^{(N)}$, is introduced in accordance with the general conception of the
transformation operators \cite{d} as an $N$-order differential operator
intertwining two Hamiltonians $h_0$ and $h_N$. The proper functions $\varphi
_E(x)$ of  one of them (for example $h_0$) are assumed to be known: $%
h_0\varphi _E(x)=E\varphi _E(x)$. One then obtains the proper function $\psi
_E(x)$, corresponding to the same eigenvalue $E$, of the other (i.e. $h_N$)
with the help of the operator $L^{(N)}$: $\psi _E=L^{(N)}\varphi _E$, $%
h_N\psi _E(x)=E\psi _E(x)$ except for the functions which form the kernel of
the operator Laplace adjoint to $L^{(N)}$ denoted by $L^{(N)^{+}}$. We
assume that the operators $h_0$ and $h_N$ are self-adjoint. (More precisely
we suppose that their potentials are real-valued functions and the
Hamiltonians are essentially self-adjoint in the sense of some scalar
product). In this case operator $L^{(N)^{+}}$ assures the transformation in
the inverse direction: from the eigenfunctions $\psi _E$ to the
eigenfunctions $\varphi _E.$ When $N=1$ we have the well-known Darboux
transformation \cite{Dar} called {\it first-order Dardoux transformation.}

It can be shown \cite{bstmf} that the operator $L^{(N)}$ can always be
presented as a product of $N$ first-order Darboux transformation operators
between every two juxtaposed Hamiltonians $h_0$, $h_1$, \ldots , $h_N$: $%
L^{(N)}=L_{0,1}L_{1,2}\ldots L_{N-1,N}$, $L_{p.p+1}h_p=h_{p+1}L_{p,p+1}$, $%
p=0,1,\ldots ,N-1$. Some of the intermediate Hamiltonians $h_p$ can have
complex-valued potentials but the final potential of the $h_N$ remains
always real-valued function (so-called irreducible case \cite{andr})

In this latter we want to point out that other than that described in Ref. 
\cite{andr} irreducible case exists. It is connected with the choice of
discrete spectrum functions of the Hamiltonian $h_0$ as the transformation
functions. In this case the intermediate potentials are real-valued
functions having additional singularities with respect to initial potential.

It follows from theorem proved in Ref. \cite{bstmf} that the operator $%
L^{(N)}$ can always be presented in the form known as Crum-Krein formula 
\cite{cr}, \cite{kr}: 
\begin{equation}
\label{e1}L^{(N)}=W^{-1}(u_1,u_2,\ldots ,u_N)\left| 
\begin{array}{cccc}
u_1 & u_2 & \cdots  & 1 \\ 
u_1^{\prime } & u_2^{\prime } & \cdots  & d/dx \\ 
\vdots  & \vdots  & \ddots  & \vdots  \\ 
u_1^{(N)} & u_2^{(N)} & \cdots  & d^N/dx^N
\end{array}
\right| 
\end{equation}
where $W$ stands for the usual symbol for the Wronskian of the functions $%
u_1,u_2,\ldots ,u_N$ called transformation functions and satisfied the
initial Schr\"odinger equation ($h_0u_i=\alpha _iu_i$), the prime denotes
the derivative with respect to real coordinate $x$, and the determinant is a
differential operator obtained by the development of the determinant in the
last column with the functional coefficients placed before the derivative
operators. Potential difference between the final Schr\"odinger equation
potential and the initial one reads as follows: $A_N(x)=V_N(x)-V_0(x)=-2[%
\log W(u_1,u_2,\ldots ,u_N)]^{\prime \prime }$.

The function $A_N(x)$ is well defined if the Wronskian $W(u_1,u_2,\ldots
,u_N)$ conserves its sign in the interval $R=[a,b]$ for the variable $x$ in
the initial Schr\"odinger equation. If the discrete spectrum eigenfunctions $%
u_i$ of the Hamiltonian $h_0$ are enumerated by the number of their zeros,
the condition for the Wronskian to conserve its sign is formulated by Krein 
\cite{kr}: the Wronskian $W(u_{k_1},u_{k_2},\ldots ,u_{k_N})$ conserves its
sign, the integers $k_i$ being equal to the number of zeros of functions $%
u_{k_i}$, if for all $k=0,1,2,\ldots $, the following inequality: $%
(k-k_1)(k-k_2)\ldots (k-k_N)\geq 0$ holds. In particular, the functions $%
u_{k_i}$ may be two-by-two juxtaposed discrete spectrum eigenfunctions. The
levels with $E=E_{k_i}$, $i=1,\ldots ,N$ will be absent in the discrete
spectrum of the new Hamiltonian $h_N$.

It follows from the formula (\ref{e1}) that $\ker L^{(N)}=$span$\{u_i\}$.
For $\ker L^{(N)^{+}}$ we have: $\ker L^{(N)^{+}}=$span$\{v_i\}$ where \cite
{bstmf} $v_k=W^{(k)}(u_1,u_2,\ldots ,$ $u_N)W^{-1}(u_1,u_2,\ldots ,u_N)$, $%
W^{(k)}(u_1,u_2,\ldots ,u_N)$ is the $N-1$-order Wronskian constructed from
the functions $u_1,u_2,\ldots ,u_N$ except for the $u_k$, $k=1,\ldots ,N$.

The product $L^{(N)^{+}}L^{(N)}$ being a symmetry operator for the initial
Schr\"odinger equation is a polynomial function of the initial Hamiltonian.
Taking into account the condition $L^{(N)}u_i=0$, $i=1,\ldots ,N$ we obtain
more precisely \cite{andr}, \cite{bstmf}:%
$$
L^{(N)^{+}}L^{(N)}=\prod\limits_{i=1}^N(h_0-\alpha _i). 
$$
The same is true for the product $L^{(N)}L^{(N)^{+}}$:%
$$
L^{(N)}L^{(N)^{+}}=\prod\limits_{i=1}^N(h_N-\alpha _i). 
$$

{\bf 3}. Let $u_i$ be two-by-two juxtaposed discrete spectrum eigenfunctions
of $h_0$ and $\{\varphi _i,\varphi _\lambda \}$ be a basis set of the
Hilbert space $H(R)$ ($\{\varphi _i\}$ is a discrete subsystem and $%
\{\varphi _\lambda \}$ is a continuous one). Introduce the notation $%
N_0=\{i:u_i=\varphi _i\}$. Then the system of functions $\{L^{(N)}\varphi _i,
$ $L^{(N)}\varphi _\lambda ,$ $i\notin N_0\}$ is complete in $H(R)$ \cite{kr}%
, \cite{ad}. With the help of $L^{(N)}$ and $L^{(N)^{+}}$ we built up the
supercharges $Q=\left( 
\begin{array}{cc}
0 & 0 \\ 
L^{(N)} & 0
\end{array}
\right) =(Q^{+})^{\dagger }$ which together with the super-Hamiltonian $%
{\cal H}=$diag$(h_0,h_N)$ form an $N$-order superalgebra \cite{andr}, \cite
{bstmf}: 
$$
[Q,{\cal H}]=[Q^{+},{\cal H}]=0,\{Q,Q^{+}\}=\prod_{i=1}^N({\cal H}-\alpha
_i). 
$$

Every energy $E=E_i$ of the superhamiltonian ${\cal H}$ is twofold
degenerate if $i\notin N_0$ and nondegenerate if $i\in N_0$. The energy $%
E_{i_0}=\min \limits_{i\in N_0}\{E_i\}$ can be associated with the ground
state of the super-Hamiltonian ${\cal H}$ and the wave function $\Psi
_{i_0}=(\varphi _{i_0},0)^T$ being annihilated by both supercharges can be
considered as the vacuum state. All the other nondegenerate states $\Psi
_i=(\varphi _i,0)^T$, $i\in N_0$ are also annihilated by both supercharges.
We can choose the set $\{u_i,i\in N_0\}$ in such a way that the ground state
of the Hamiltonian $h_0$ does not belong to this set. In this case proper
functions of the super-Hamiltonian with the energies below to its vacuum
state exist and are twofold degenerate.

4. We will cite an example of the above-described situation. Consider the
harmonic potential $h_0=-d^2/dx^2+x^2/4-1/2$ with the discrete spectrum $%
E_n=n=0,1,2,\ldots $ and the well-known discrete spectrum eigenfunctions $%
\varphi _n=(\sqrt{2\pi }n!)^{-1/2}\times \\\exp (-x^2/4)He_n(x)$ where $%
He_n(x)$ is the Hermit polynomial \cite{abr}. The double Darboux
transformation with the juxtaposed functions $\varphi _k$ and $\varphi _{k+1}
$, $k\geq 0$ produces a new potential of the form \cite{bstmf}:%
$$
\begin{array}{c}
V_2(x)=
\frac{x^2}4+\frac 32-2\frac{J_k^{\prime \prime }(x)}{J_k(x)}+2\left( \frac{%
J_k^{\prime }(x)}{J_k(x)}\right) ^2, \\ J_k(x)=\sum\limits_{i=0}^k\frac{%
\Gamma (k+1)}{\Gamma (i+1)}He_i^2(x).
\end{array}
$$
In its discrete spectrum the levels $E=k$ and $E=k+1$ are absent. The
normalized to unity wave functions have the form%
$$
\begin{array}{c}
\psi _n(x)=[
\sqrt{2\pi }n!(n-k)(n-k-1)]^{-1/2}\exp (-x^2/4) \\ \times
[(n-k)He_n(x)+f_{kn}(x)
\frac{He_{k+1}(x)}{J_k(x)}], \\ 
f_{kn}(x)=He_k(x)He_{n+1}(x)-He_n(x)He_{k+1}(x),\quad n\neq k,(k+1).
\end{array}
$$
The state $\Psi _0=(\varphi _k,0)^T$ having the minimal energy value among
the two states annihilated by both supercharges can be associated with the
vacuum state. All the states $\Psi _{i-k}=a(\varphi _i,0)^T+b(0,\psi _i)^T$, 
$i<k$, $a,b\in {\Bbb C}^1$ have the energies below  energy $E=k$ of the
vacuum state $\Psi _0$.

{\bf 5.} The supersymmetric quantum mechanics is now widely used in
different branches of physics such as statistical physics, condensed matter,
atomic physics \cite{Coop}, \cite{R1}. An essential ingredient of this
theory is the Darboux transformation which permits us to construct for every
exactly solvable potential a family of its exactly solvable partners. If we
start (as in our example) from the harmonic potential we can construct
exactly solvable potentials with equidistant or quaziequidistant spectra 
\cite{bstmf}. Coherent states of these potentials \cite{R2} known in quantum
optics as wavelets being constructed represent nondispersive wave packets.

The above considered potential in the particular case $k=1$ has been
obtained earlier by other means \cite{R3}. This potential corresponds to the
one of the rational solutions of the Painlev\'e IV differential equation 
\cite{R4}. The connection of the Painlev\'e IV and V transcendents with the
Schr\"odinger equation was studied in Ref. \cite{R5}. In our opinion with
the help of the Darboux transformation it is possible to establish the
correspondence between the known rational solutions of these equations \cite
{R6} and exactly solvable potentials with the quaziequidistant spectra.

An application of the double Darboux transformation to the Coulomb potential
which gives a new exactly solvable potential, in which the discrete spectrum
the levels with $n=2$ and $n=3$ are absent, was made in Ref. \cite{R7}.
Using these results and the above described approach we can construct for
this system the second-order supersymmetric model analogous to the one
discussed here.

\end{document}